\documentclass[12pt,aps]{revtex4}
\pdfoutput=1

\newcommand{\beq}{\begin{equation}}
\newcommand{\eeq}{\end{equation}}

\usepackage{graphicx}
\usepackage{color}

\begin{document}

\title{Electric field-dependent dynamic polarizability and state-insensitive conditions for optical trapping of 
diatomic polar molecules} 

\author{Svetlana Kotochigova$^{1*}$ and David DeMille$^2$}

\vspace*{0.5cm}

\affiliation{$^1$Department of Physics, Temple University, Philadelphia, PA 19122-6082, USA\\
$^2$Department of Physics, Yale University, New Haven, CT 06520, USA}

\begin{abstract}
Selection of  state-insensitive or ``magic'' trapping conditions with ultracold atoms or
molecules, where pairs of internal states experience identical trapping
potentials, brings  substantial benefits to precision measurements and
quantum computing schemes. Working at such conditions could ensure that
detrimental effects of inevitable inhomogeneities across an ultracold
sample are significantly reduced. However, this aspect of confinement
remains unexplored for ultracold polar molecules. Here, we present means
to control the AC Stark shift of rotational states of ultracold diatomic polar
molecules, when subjected to both trapping laser light and an external
electric field.  We show that both the strength and relative orientation
of the two fields influence the trapping potential.  In particular,
we predict  ``magic electric field strengths" and a ``magic angle",
where the Stark shift is independent  of the DC external field for certain
rotational states of the molecule.  
\end{abstract}

\maketitle

The advantage of using state-independent light traps for precision
frequency measurements with ultracold atoms has been demonstrated
in several experiments \cite{Katori,TrapScience08}.  Applications of
this approach were analyzed in the context of optical atomic clocks
and coherent control of atoms and photons within an optical cavity. For
these applications the frequency of the laser beam that creates a far-off
resonant optical dipole trap is chosen such that the AC Stark shift of
the ground and one excited electronic  atomic level are the same. In this
way any optical atomic  transition between these levels is unaffected by
the trapping light.  In Refs. ~\cite{Flambaum,Beloy,Derevianko,Lundblad}
state-insensitive trapping conditions have also been found for microwave
transitions in the atomic ground state by using a combination of the
vector and tensor components of the AC Stark shift and an external
magnetic field.

Ultracold molecular systems possess unique proporties that are
considered to make them potentially useful as tools for precision
measurements \cite{special_issue,NJP_review} and quantum computing
\cite{DeMille02,Zhao}. Therefore it is desirable to extend the
zero-differential AC Stark shift technique to these more complex
systems.  The idea of ``magic''  frequencies for vibrational Raman
transitions in homo-nuclear Sr$_2$ molecules was first explored in
Refs.~\cite{Zelevinsky,Kotochigova09}. This molecule  is proposed for
a search of possible time variation of the electron-to-proton mass ratio.
The AC Stark shift of a molecule  is determined by the dynamic molecular
polarizability $\alpha(\nu)$, which is a function of radiation frequency $\nu$
and its polarization.

Polar molecules have a permanent dipole moment and their levels can
be shifted and mixed with one another by applying an external electric
field. This opens up a new way to create ``magic''  trapping conditions
for two rotational levels of the molecule. 
In the presence of an external electric
field, $J$ is not a good quantum number and all states, even the ``rotationless''
ground state, have an anisotropic polarizability~\cite{Ospelkaus}.
The anisotropy of the dynamic polarizability
of these levels manifests itself as a dependence  on the relative
orientation of the polarization of the trapping laser and the DC electric
field.  The combined action of these two fields can be
a powerful tool to manipulate and control ultracold molecules trapped
in an optical potential.

The behavior described here has potential applications to several
experiments that have been envisioned for diatomic polar molecules held in
optical traps.  The implications are particularly striking for the
use of polar molecules in an optical lattice as quantum bits. As
was described in Ref. \cite{DeMille02}, a pair of rotational states forms a 
suitable quantum bit.  However,  it was pointed out that this
system is susceptible to decoherence due to intensity fluctuations in
the optical trapping lasers, if the dynamic polarizabilities of these
levels differ as is generally the case.  This in turn leads to very
stringent requirements on the laser intensity stability for such a
system to be practical.  As we will show, this limitation can
be removed by adjusting the experimental parameters to guarantee that
the dynamic polarizabilities of these states are equal.  In particular,
for this proposed system (where a spatial gradient of the electric field
$\vec{\mathcal{E}}$ is required, so working at a magic electric field
value is impossible) it should be possible to use light polarized at the
``magic angle'' relative to the static field in order to eliminate this
potentially dangerous source of decoherence.

The existence of magic electric field values is also of possible use
for envisioned applications where polar molecules in optical lattices
are employed in novel types of many-body systems.  This includes,
for example, cases where the properties of long-range molecular
interactions are tailored by a combination of a static electric field
and resonant microwave fields coupling different rotational states
\cite{Micheli06,Buchler06, Buchler07}.  Working at a magic electric field
value in such systems could ensure that the inevitable inhomogeneities
in the intensity of the trapping light across a large sample would not
change the resonant condition for the microwave drive fields.  Hence,
working under such ``magic'' conditions might be necessary to implement
proposals of this type.

Motivated by these ideas, we calculate  the near-infrared dynamic
polarizability of various rotational levels of the $v=0$ vibrational
state of the X$^1\Sigma^+$ potential of the KRb and RbCs molecules,
under the simultaneous influence of trapping electromagnetic and
static electric fields.  We calculate dynamic polarizability using
the computational techniques developed in previous publications
\cite{Kotochigova1,Kotochigova2,Kotochigova3}.  Our calculations with an
external electric field are predominantly performed at a laser frequency
of 9174 cm$^{-1}$ (or wavelength of 1090 nm), which corresponds to an
often-used frequency to trap atoms and molecules in ultracold experiments.
In the near infrared this laser frequency is sufficiently far away
from molecular resonances of the excited electronic states that heating
due to photon scattering is negligible.  We focus on external electric
field strengths  up to 15 kV/cm, a value which should  be experimentally
accessible. \\

\begin{figure}
\includegraphics[scale=0.3]{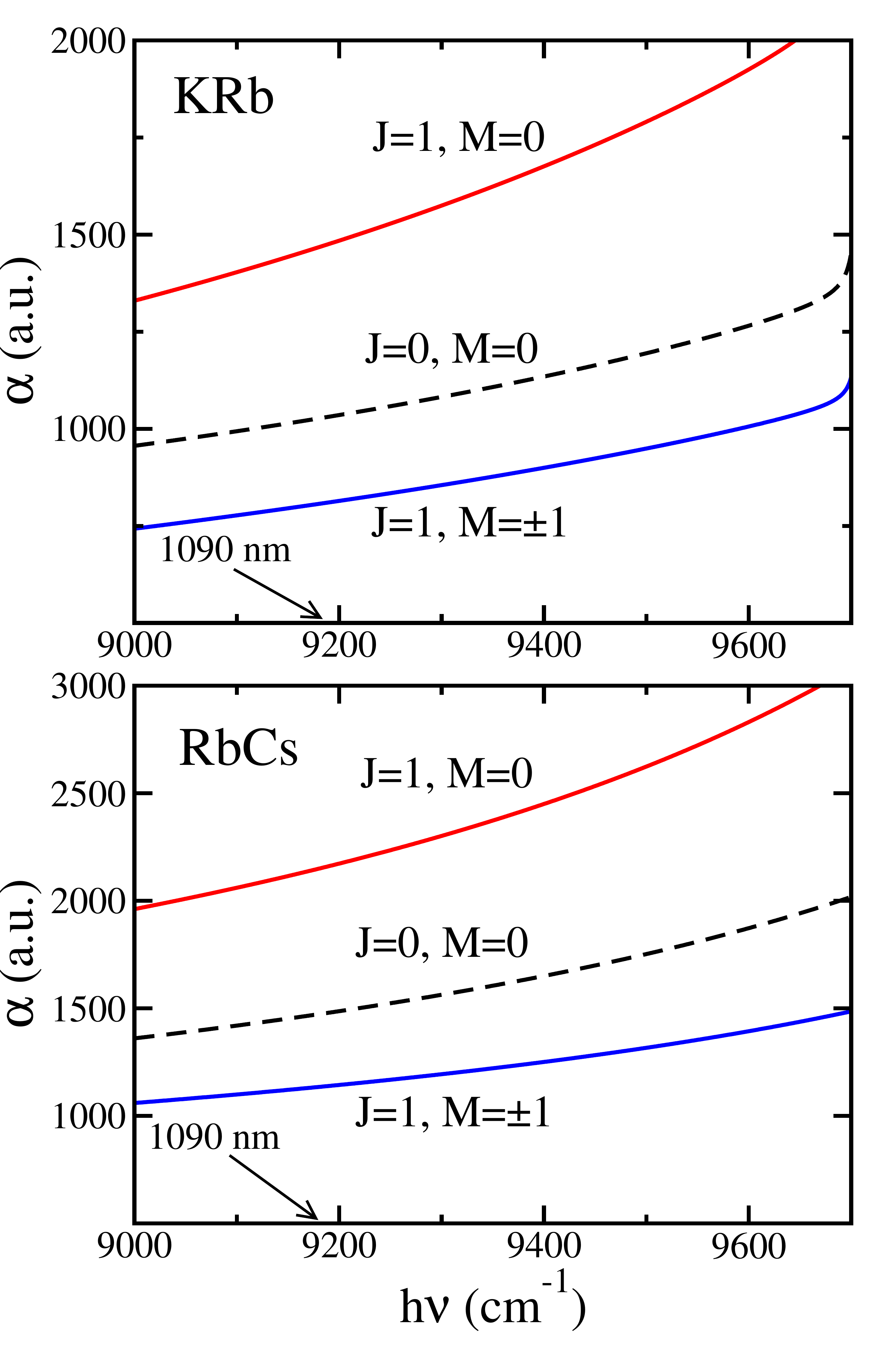}
\caption{Dynamic polarizability in the absence of an electric field and
in atomic units for the $v$=0, $J$=0 and 1 levels of the X$^1\Sigma^+$
ground state of KRb and RbCs as a function
of trapping frequency in the near infrared domain. The dashed lines show
the polarizability for the $J$=0 and $M$ = 0 level, which is independent
of the light polarization.  The solid lines correspond to the $J$= 1
state with magnetic sublevels $M$ = 0 and $\pm$ 1 illuminated by linear
polarized light along the $\hat z$ direction. One atomic unit of polarizability corresponds to
4.68645$\times 10^{-8}$ MHz/(W/cm$^2$). } 
\label{zero-field} 
\end{figure}

We begin by studying the dynamic polarizability or AC Stark shift for the
$J=0$ and $J=1$ rotational levels  of the $v=0$ X$^1\Sigma^+$ state
of KRb and RbCs in the near infrared, without an external electric
field. The  procedure used to determine the complex molecular
polarizability has already been described in our previous paper
\cite{Kotochigova1}. Figure~\ref{zero-field} illustrates the results of
this study in the absence of an electric field by showing the dynamic polarizability $\alpha$
as a function of laser frequency $h\nu$. The range of laser frequencies
spans a technologically relevant near-infrared optical domain and includes 
the lasing frequency near 9174 cm$^{-1}$ used in Ref. \cite{Ospelkaus}.
The figure also shows that the dynamic polarizability of the molecular state depends on the
rotational quantum number $J$ and its projection $M$. The curves for the different states
do not cross as a function of the laser frequency.   In other words, at zero electric
field we can not find a ``magic'' frequency for J=0 and J=1 states using
trapping light in the near infrared.\\

\noindent
{\bf Field-dependent dynamic polarizability}\\

\noindent
Here we extend the idea of the AC Stark shift for the rotational levels
of the  X$^1\Sigma^+$ ground state to the mixing of these levels in a static electric
field $\vec{\mathcal{E}} = \mathcal{E}\hat{z}$ along the space-fixed $\hat{z}$ direction. 
Our analyses show that the experimentally accessible electric fields mix only a few 
low-lying rotational states. The ultimate goal is to introduce the dynamic polarizability
of the mixed rotational levels as a function of an external DC electric field and the polarization
of the AC trapping field.
In a heteronuclear molecule this Stark mixing is primarily due to the permanent electronic
dipole moment $\vec d$; we neglect the much smaller effects due to
Stark mixing with other electronic states.  For the Hund's case
(a) X$^1\Sigma^+$  state of an alkali-metal dimer, the total electron spin
and orbital angular momentum are not coupled to the molecular rotation.
The molecular wavefunction in the lab frame, $| vJM\rangle_z$, is then given by 
\begin{equation}
         | vJM\rangle_z \equiv |vJ\rangle \times Y_{JM}(\hat R) = 
         \left\{ \frac{ \psi_{vJ}(R)}{R}| {\rm X}^1\Sigma^+ \rangle \right\} 
         \times Y_{JM}(\hat R)\,  \,,
\end{equation} 
where $v$ is a vibrational quantum number,  $J$ and $M$
are the molecular rotational angular momentum and its projection along
the $z$ axis, $\psi_{vJ}(R)$ is the radial rovibrational wavefunction,
$Y_{JM}(\hat R)$ is a spherical harmonic, $ \hat R$ is the orientation
of the molecule relative to the electric field direction $z$, $| {\rm
X}^{1}\Sigma^+  \rangle$ is the electronic wavefunction with
 projections defined along the internuclear axis.
For the X$^1\Sigma^+$ state, in even modest electric fields the nuclear spins are well-decoupled from the
other spins and angular momenta; hence we ignore the nuclear spins here.  Then the Hamiltonian for such a system
becomes 
\begin{equation}
   H =  \sum_{vJM} E_{vJ} | v JM\rangle_z
                       \times    {}_z\langle v JM |   - \vec d \cdot
                       \vec{\mathcal{E}} \,,
\label{HE} 
\end{equation} 
where $E_{vJ}= G_v + B_v J(J+1)$ and $G_v$ and $B_v$ are the
vibrational energy and rotational constant of vibrational level $v$,
respectively. Higher order rotational corrections are negligible.  For the
alkali-metal dimers KRb and RbCs we have that $\Delta G_v=G_{v+1}-G_v$
is on the order of 50-100 cm$^{-1}$ for small $v$, while $B_v$ is on
the order of 0.017-0.037 cm$^{-1}$.

We evaluate the matrix elements of the operator $-\vec d \cdot \vec{\mathcal{E}}$
by noting  that after averaging over the electronic wavefunction $|
{\rm X}^1\Sigma^+ \rangle$ it reduces to $-d(R) C_{10}(\hat R) \mathcal{E}_z$, 
where $d(R)= \langle{\rm X}^1\Sigma^+ |d| {\rm X}^1\Sigma^+ \rangle$ is the
$R$-dependent permanent electric dipole moment, 
$C_{lm}(\hat R)=\sqrt{4\pi/(2l+1)}Y_{lm}(\hat R) $ 
are tensors of rank $l$,
and $\mathcal{E}_z$ is the electric
field strength.  
Consequently, this operator conserves the projection quantum number $M$.  The matrix element between two rovibrational states is
\begin{eqnarray}
\lefteqn{_z\langle vJM | -\vec d \cdot \vec{\mathcal{E}}  | v'J'M'\rangle_z =- \delta_{MM'}\,
d_{vJ,v'J'} \, \mathcal{E}_z\int d \hat R \,Y_{JM}^*(\hat R) C_{10}(\hat R) Y_{J'M'}(\hat R)} \label{DM}\\
                   &  & \quad\quad =   - \delta_{MM'}   d_{vJ,v'J'}\mathcal{E}_z
         (-1)^M   \sqrt{(2J+1)(2J'+1)} 
            \left( \begin{array}{ccc} J & 1 & J' \\ -M& 0 & M' \end{array} \right)
                        \left( \begin{array}{ccc} J & 1 & J' \\ 0& 0 & 0 \end{array} \right) \,,
 \nonumber                       
\end{eqnarray}
where $\delta_{MM'}$ is the Kronecker delta function, $d_{vJ,v'J'}=  \int_0^\infty dR\, \psi_{vJ}(R) d(R) \psi_{v'J'}(R)$, 
and $(\cdots)$ are 3-$j$ symbols (see e.g. \cite{Brink}). 
This matrix element is nonzero when $J+1+J'$ is even, according to the parity selection rules, and is 
independent of the sign of $M$ and $M'$.
For the small $J$ values of interest here we can assume that the $J$ dependence of $\psi_{vJ}(R)$
is negligible.
Moreover, coupling between vibrational levels  can also be ignored, as the dipole moment $d(R)$ is a slowly
varying function with $R$ and the spacing between vibrational levels is large  compared to the rotational splitting. 
For tensor operators $C_{lm}(\hat R)$  of rank 0, 1,  and 2 the relationship between Cartesian $x$, $y$, and $z$ 
components and spherical $m=-1$, $0$, and $+1$ components  can be found in Refs.~\cite{Bonin,Varsholovich}.

For each projection $M$ and vibrational level $v$, eigenvalues and eigenvectors of the
Hamiltonian, Eq.~(\ref{HE}), are obtained by the direct diagonalization of the Hamiltonian
matrix  including  rotational states $J$ from $|M|$ up to some value $J_{\rm max}$. For
the external electric field strength and $J$ values of interest, $J_{\rm max}=10$ is
sufficient for convergence.  We label the eigenenergies by $E_{v\tilde J M}$ with
corresponding eigenvectors $|v\tilde J M \rangle = \sum_{J}U^{vM}_{\tilde J,J}|vJM\rangle_z$.
Here $\tilde J$ is an integer index with values $\tilde J = |M|, |M|+1,...$, such that the
eigenstate $|v\tilde J M \rangle$ adiabatically connects to the electric field-free
eigenstate $|vJM\rangle_z$ with $J = \tilde J$.  
For $|M|>0$ the levels with projection quantum number $-M$ and $M$ remain degenerate.
The dipole matrix element
between states $|v\tilde J M\rangle$ and $|v' \tilde J' M'\rangle$ of the X potential is given
by $ \langle v\tilde J M | d_{\sigma} | v' \tilde J' M'\rangle = \sum_{J,J'} U^{vM}_{\tilde J,J}
\,U^{v'M'}_{\tilde J',J'} \times {}_z\langle vJM | d_{\sigma} | v'J'M'\rangle_z$, where
$\sigma$ is a spatial index that can be expressed either in Cartesian coordinates $x,y,z$,
or spherical coodinates $q = 0,\pm 1$.

Now we are able to calculate the dynamic polarizability of the mixed rotational 
eigenstates at laser frequency $\nu$.  It is determined by the properties of the operator $\alpha_{\sigma\sigma'}(\nu)$ defined by
\cite{Bonin,Stone}
\begin{eqnarray}
            \alpha_{\sigma \sigma'}(\nu)
            & =& \sum_{\gamma}
                        \left \{ \frac{1}{E_{\gamma} - E_{v\tilde J M} + h\nu}
                              + \frac{1}{E_{\gamma} - E_{v\tilde J M} - h\nu} \right \}
                         d_{\sigma} | \gamma \rangle \langle \gamma | d_{\sigma'} 
\label{operator_def}
\end{eqnarray}
with $\gamma$ enumerating  eigenstates of the ground as well as excited
electronic potentials in the presence of the electric field and $\sigma,\sigma'=x,y,$ or $z$.  
We will focus on the $J=0$ and $J=1$ levels and without loss of generality assume that the $x$ axis of our coordinate system lies in the plane spanned by the electric-field direction and the orientation of the linearly-polarized laser light. Moreover, we are interested in the situation where the level
shifts due to the laser are small compared to those induced by the electric field. For the isolated levels with M=0, the polarizability is determined by the diagonal matrix element  $\alpha^{v\tilde{J}M=0}_{\sigma\sigma'}(\nu) = \langle v\tilde{J}M=0 | \alpha_{\sigma\sigma'}(\nu) | v\tilde{J}M=0 \rangle$.  The $|M|>0$ dynamic polarizability needs to be treated by degenerate perturbation theory within the two-dimensional subspace $|v\tilde J M\rangle$ and $|v\tilde J -\!M\rangle$.
In fact,  our choice of the $x$ direction and the symmetry properties of  the dynamic polarizability ensure that the linear combinations
$|v\tilde J M,\pm\rangle=\{|v\tilde J M\rangle\pm|v\tilde J -\!M\rangle\}/\sqrt{2}$ with $M>0$ are the correct eigenstates.
For these states  the dynamic polarizability $ \alpha_{\sigma\sigma'}^{v\tilde J M,\pm}(\nu)= 
\langle v\tilde J M, \pm  |  \alpha_{\sigma \sigma'}(\nu)|v\tilde J M, \pm\rangle$.
For diatomic species, 
only the diagonal elements $\alpha^{v\tilde J M,\pm}_{xx}$, $\alpha^{v\tilde J M,\pm}_{yy}$, and $ \alpha^{v\tilde J M,\pm}_{zz}$  are nonzero. 
Hence in an oscillating electric field $\vec{\mathcal{E}}_o(t)= \mathcal{E}_o(0) \textrm{Re}\{\vec{\epsilon} e^{i 2\pi \nu t}\}$ (where $\vec{\epsilon}$ is the complex unit vector indicating
the polarization), the state $|v\tilde J M,\pm\rangle$ shifts in energy by an amount $\Delta E$ given by
$\Delta E = -\sum_{\sigma,\sigma'}|\mathcal{E}_o(0)|^2 \alpha^{v \tilde J M,\pm}_{\sigma \sigma'}(\nu)\epsilon_{\sigma}\epsilon^*_{\sigma'}/4$. The sum is over the spatial indices $\sigma,\sigma'=x,y,z$.

We are interested in near infrared laser frequencies, which are detuned
away from resonances with rovibrational levels of the electronically
excited potentials.  In particular, we focus on wavelengths between
1000 nm and 1100 nm. Starting from the $v=0$ vibrational level
of the X$^1\Sigma^+$ states of KRb and RbCs, photons of this wavelength 
do not possess sufficient energy to reach
rovibrational levels of the electronically excited singlet $^1\Lambda$ potentials. 
The photons do have enough energy to reach vibrational level of the triplet b$^3\Pi$
potential considering only single-photon excitations. However, such transitions 
require relativistic spin-orbit coupling to
the A$^1\Sigma^+$ potential to acquire a non-zero dipole matrix element; such spin-orbit 
induced couplings are small enough to neglect under the conditions of interest here. 

This allows us to make several approximations.  Firstly, we can
use non-relativistic potentials and transition dipole moments and in
the calculation of the dynamic polarizability only consider singlet
$^1\Sigma^+$ and $^1\Pi$ potentials. Secondly, assuming a large detuning (such that 
$|E_\gamma - E_{v\tilde J M} - h\nu | \gg E_{vJM}$ for all states of interest)
we can neglect the electric field  and rotational dependence of the
energy denominators in Eq.~(\ref{operator_def}).  Finally, we find that
in the near infrared the contribution to the polarizability from intermediate states $\gamma$ in the
ground X$^1\Sigma^+$ state is small.

With this in mind the polarizability becomes
\begin{eqnarray}     
 \alpha^{v\tilde J M,\pm}_{\sigma \sigma'}(\nu) & \cong &   
   \sum_{J,J'}        U^{v|M|}_{\tilde J,J} U^{v|M|}_{\tilde J,J'}   \sum_{e v_e \Lambda}
                       \left \{ \frac{1}{E_{ev_e\Lambda}-E_{v}+h\nu}
                              + \frac{1}{E_{ev_e\Lambda}-E_{v}-h\nu}  \right\}    \label{excited}  \\
                              && \quad\quad \quad\quad \times
         \sum_{J_eM_e}    {}_z\langle v JM,\pm | d_\sigma | e v_e J_eM_e\Lambda\rangle_z
                  \,{}_z\langle e v_e J_eM_e \Lambda |  d_{\sigma'} | v J' M,\pm \rangle_z\,,
             \nonumber
\end{eqnarray}      
where the energy $E_{v}$ is a typical vibrational energy in the ground state potential, 
and the rovibrational wavefunctions $| e v_e J_e M_e \Lambda\rangle$ of the electronically
excited states with approximate energy $E_{ev_e\Lambda}$ are given by
\[
 | e v_e J_e M_e \Lambda\rangle_z  \equiv | e v_e \Lambda\rangle \times | J_e M_e \Lambda\rangle_z
 \equiv
     \left\{ \frac{\phi_{v_e}(R)}{R}|e~ ^1\Lambda\rangle \right\}\, \times 
         \left\{\sqrt{\frac{2J+1}{4\pi}}D^{J_e*}_{M_e\Lambda}( \hat R ) \right\}\,,
 \]
 where the vibrational and electronic dependence has been isolated
 in the ket $|e v_e \Lambda\rangle$ and the rotational dependence in $|J_e M_e \Lambda\rangle_z$, respectively.
The wavefunction $\phi_{v_e}(R)$ is the radial rovibrational wavefunction, 
the ket $|e~ ^1\Lambda\rangle$ is the electronic
state with projection quantum number $\Lambda$ defined along the
internuclear axis, and $D^J_{MM'}(\hat R)$ is a Wigner rotation matrix that describes a symmetric
top rotational wavefunction.  The sum over $\Lambda$ includes both positive and
negative values, where $\Lambda=0$ corresponds to excited  $^1\Sigma^+$
electronic states and $\Lambda=\pm 1$ to $^1\Pi$ states.  

As stated before the energy of the excited state  and
$\phi_{v_e}(R)$ depend on the electronic state $|e ~^1\Lambda\rangle$
and vibrational level $v_e$, but not  $J_e$ and $M_e$. Consequently, the transition dipole moments separate into
${}_z\langle v JM,\pm | d_\sigma | e v_e J_eM_e\Lambda\rangle_z =
                 \langle vJ | d(R) | e v_e\Lambda \rangle  F^{J_eM_e\Lambda}_{JM\pm,\sigma}$,
where $F^{J_eM_e\Lambda}_{JM\pm,\sigma}$ is an integral of the product of three Wigner rotation matrices 
$D^J_{MM'}(\hat R)$ over the orientation of the molecule $\hat R$ that can be evaluated using
 angular momentum algebra \cite{Brink}. 

Moreover, we have verified that  $ \langle vJ | d(R) | e v_e\Lambda \rangle$ is nearly independent of $J$.
The sums over $J_e$ and $M_e$ in Eq.~(\ref{excited}) can now be performed and 
 we finally find
\begin{eqnarray}        
   \lefteqn{    \langle v \tilde J M, \pm |\alpha_{\sigma \sigma'}(\nu)|v \tilde J M, \pm\rangle=  }
      \label{final} \\ 
     &&\sum_\Lambda  \alpha^\Lambda(\nu)     \sum_{J,J'}        U^{v|M|}_{\tilde J,J} U^{v|M|}_{\tilde J,J'}  
        \times \sum_{J_eM_e}
               F^{J_eM_e\Lambda}_{JM\pm,\sigma} F^{J_eM_e\Lambda}_{J'M\pm,\sigma'}
                     \nonumber
\end{eqnarray}  
with  the $\Lambda$-dependent
\begin{eqnarray}        
            \alpha^\Lambda(\nu) = 
                       \sum_{e v_e} \langle vJ | d | e v_e \Lambda \rangle
                                     \langle e v_e \Lambda | d | vJ \rangle
                         \left\{ \frac{1}{E_{ev_e\Lambda}-E_{v}+h\nu}
                              + \frac{1}{E_{ev_e\Lambda}-E_{v}-h\nu}  \right\} \,.
\end{eqnarray} 
The parallel $\alpha^0(\nu)$ and perpendicular $\alpha^1(\nu)$ contributions to the polarizability  are due to transitions to the $\Sigma$ and $\Pi$ states, respectively.\\

\noindent
{\bf Magic DC Electric Field}\\ 

\begin{figure}
\includegraphics[scale=0.3]{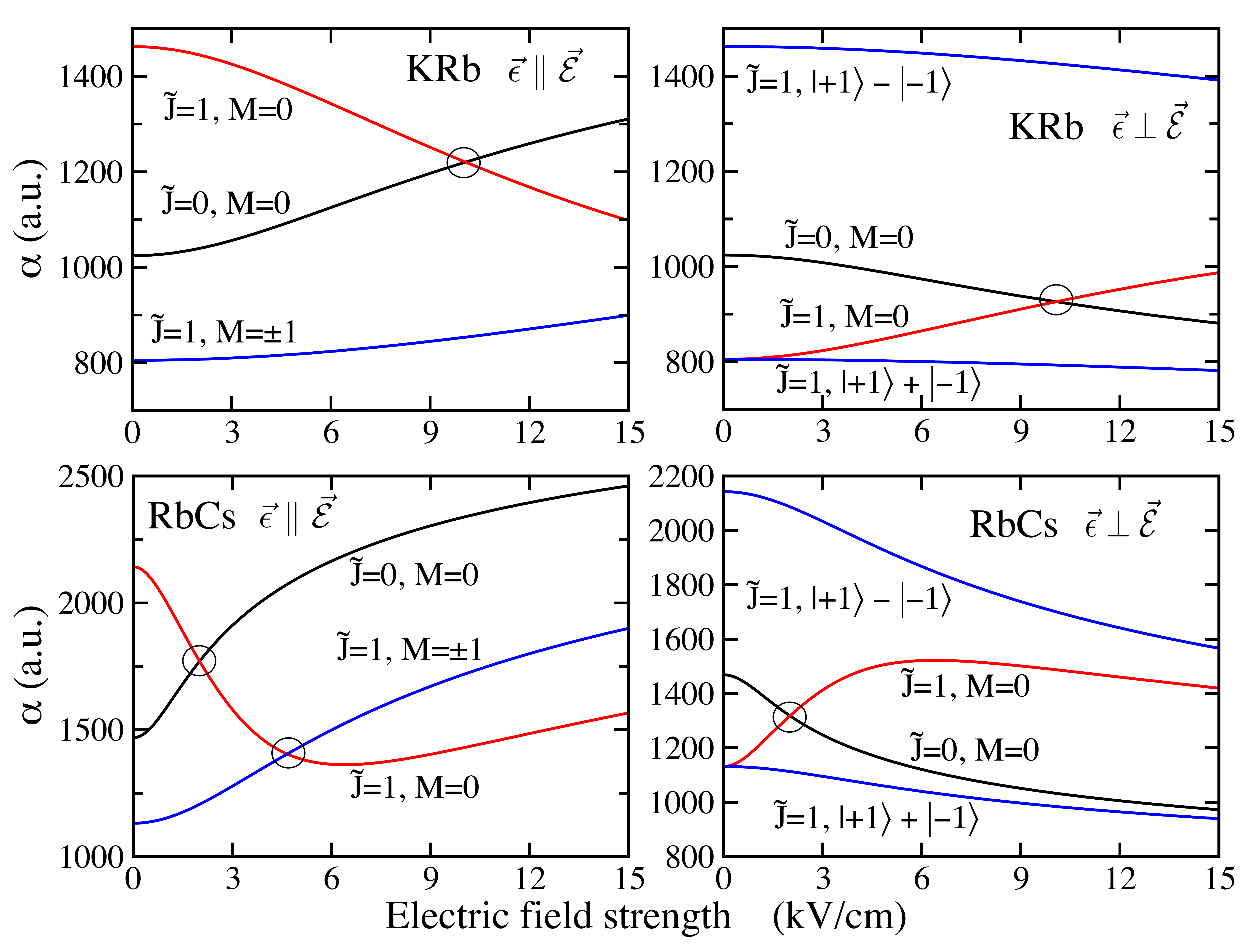}
\caption{Dynamic polarizability at a wavenumber of 9174 cm$^{-1}$ of the $v$=0, $\tilde J$=0 and 1 levels of the X$^1\Sigma^+$ 
ground state of KRb (top row) and RbCs (bottom row) as a function of the field strength of an external electric field.
The polarization of the trapping field is parallel 
(left panel) and perpendicular (right panel) to the direction of the electric 
field. The circles indicate a crossing point between $\tilde J$=0 and 1 polarizabilities.
The degeneracy of the states $\tilde J=1$ and $M=\pm 1$ is lifted by perpendicularly
polarized laser light. The new states are labeled $|+1\rangle\pm|-1\rangle$, corresponding to $|\tilde{J}=1, 
M\pm\rangle$ respectively. The  $M=\pm 1$ states remain degenerate for parallel polarization.
} 
\label{KRb_RbCs_field_orientation}
\end{figure}

The polarizability depends on the stength of an
external electric field through the use of the unitary matrices $U^{vM}_{\tilde J,J}$ 
of Eq.~(\ref{final}).   Figure~\ref{KRb_RbCs_field_orientation} 
shows the dynamic polarizability of
the ground states of the KRb and RbCs molecules as a
function of the external electric field strength. The left panels of
these figures correspond to the parallel $(\vec{\epsilon} = \hat{z})$
and the right panels to the perpendicular $(\vec{\epsilon} = \hat{x})$
polarization of the trapping light relative to the electic field direction
$\vec{\mathcal{E}}$.  In all cases the polarizability depends on both the
state index $\tilde J$ and the angular momentum projection $M$.  For KRb
(Fig.~\ref{KRb_RbCs_field_orientation}, top row) a ``magic'' electric field strength
exists at $\mathcal{E}=10$ kV/cm, where the polarizability of $\tilde J$=0,
$M$=0 and $\tilde J$=1, $M$=0 states coincide. This is possible due to the polar
character of the molecules. 
For RbCs (Fig.~\ref{KRb_RbCs_field_orientation}, bottom row)
two ``magic'' electric field strengths exist.  For the pair of
states $\tilde J=0,M=0$ and $\tilde J=1,M=0$ a crossing occurs near 2 kV/cm. Another crossing
appears for the 
states $\tilde J=1,M=0$ and $\tilde J=1,M=\pm 1$ at 4.7 kV/cm. They both are at a much smaller field
strength than for KRb, since RbCs has a larger permanent dipole
moment and smaller rotational splittings in the ground state.  Note that
the $M=0$ ``magic'' electric field occurs  at the same field
strength for both parallel and perpendicular polarization of the trapping
light. In fact, they are the same for any polarization.\\

\noindent
{\bf Magic Angle}\\

\begin{figure}
\includegraphics[scale=1]{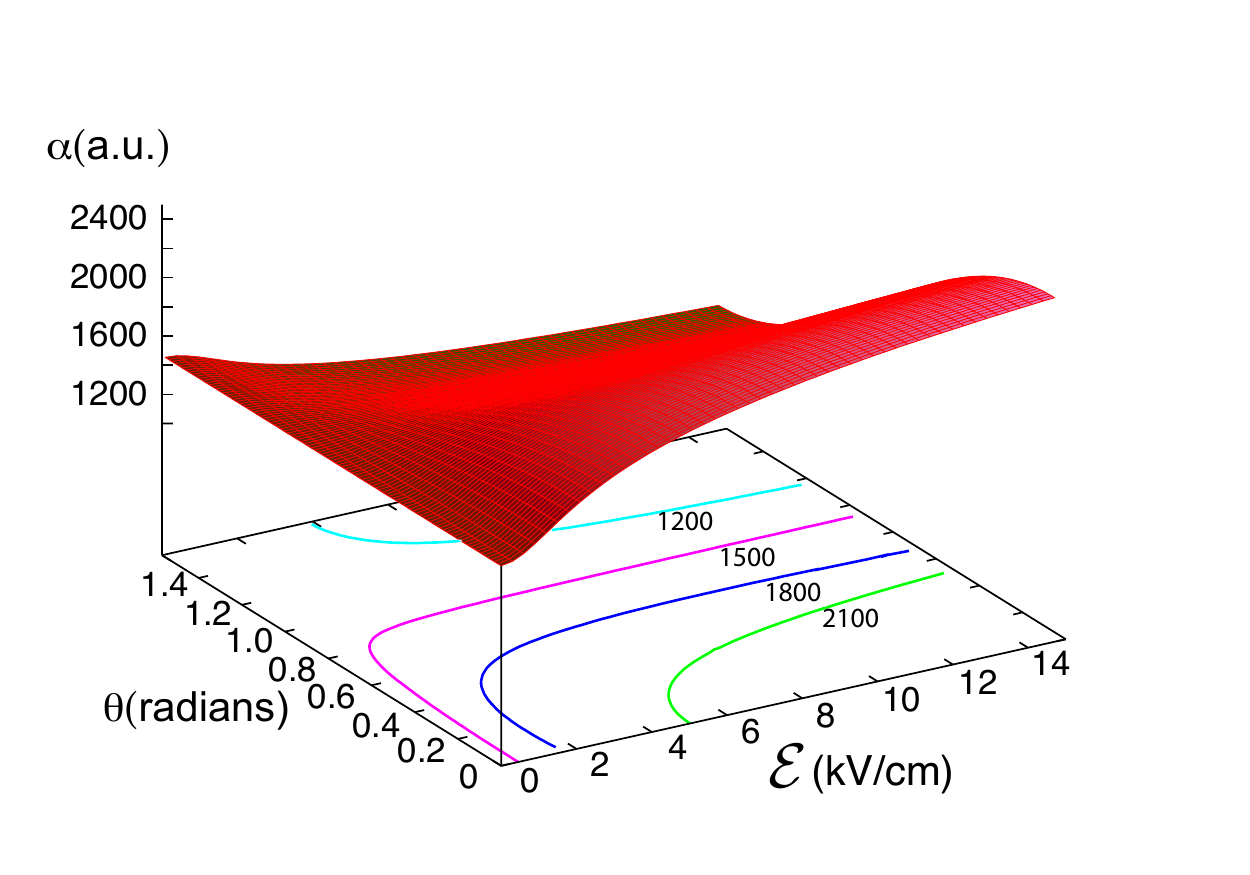}
\caption{Dynamic polarizability of the $v$=0, $\tilde J$=0, $M$= 0 level of the
X$^1\Sigma^+$ ground state of RbCs as a function of external electric field 
strength and the angle $\theta$ between the direction of the electric field $\vec{\mathcal{E}}$ and the 
polarization $\vec{\epsilon}$ of the trapping light at a wavenumber of 9174 cm$^{-1}$.
The contour lines are marked by the polarizability value in atomic units.} 
\label{E_vs_theta} 
\end{figure}

\begin{figure}
\includegraphics[scale=0.25]{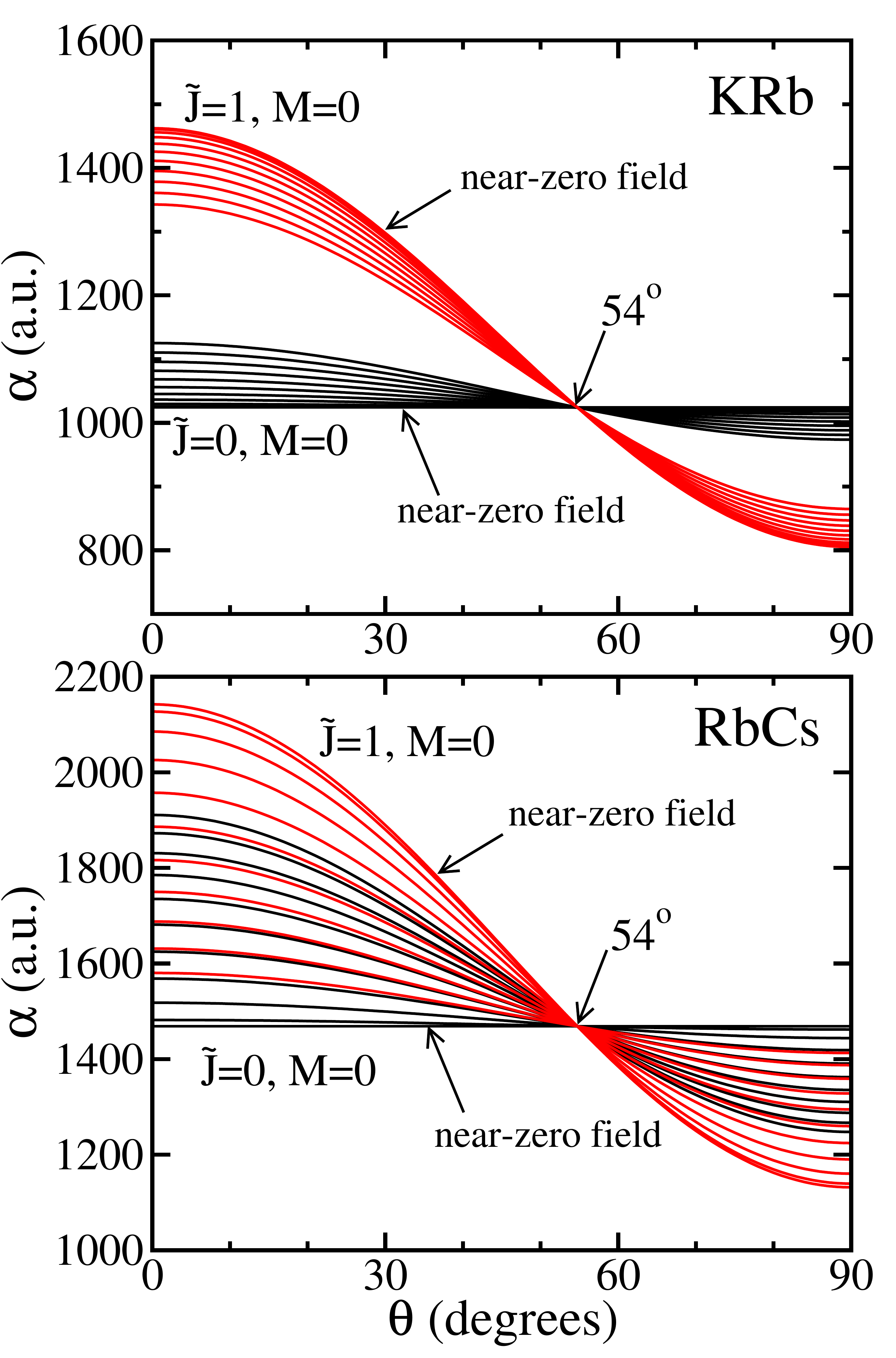}\ \includegraphics[scale=0.25]{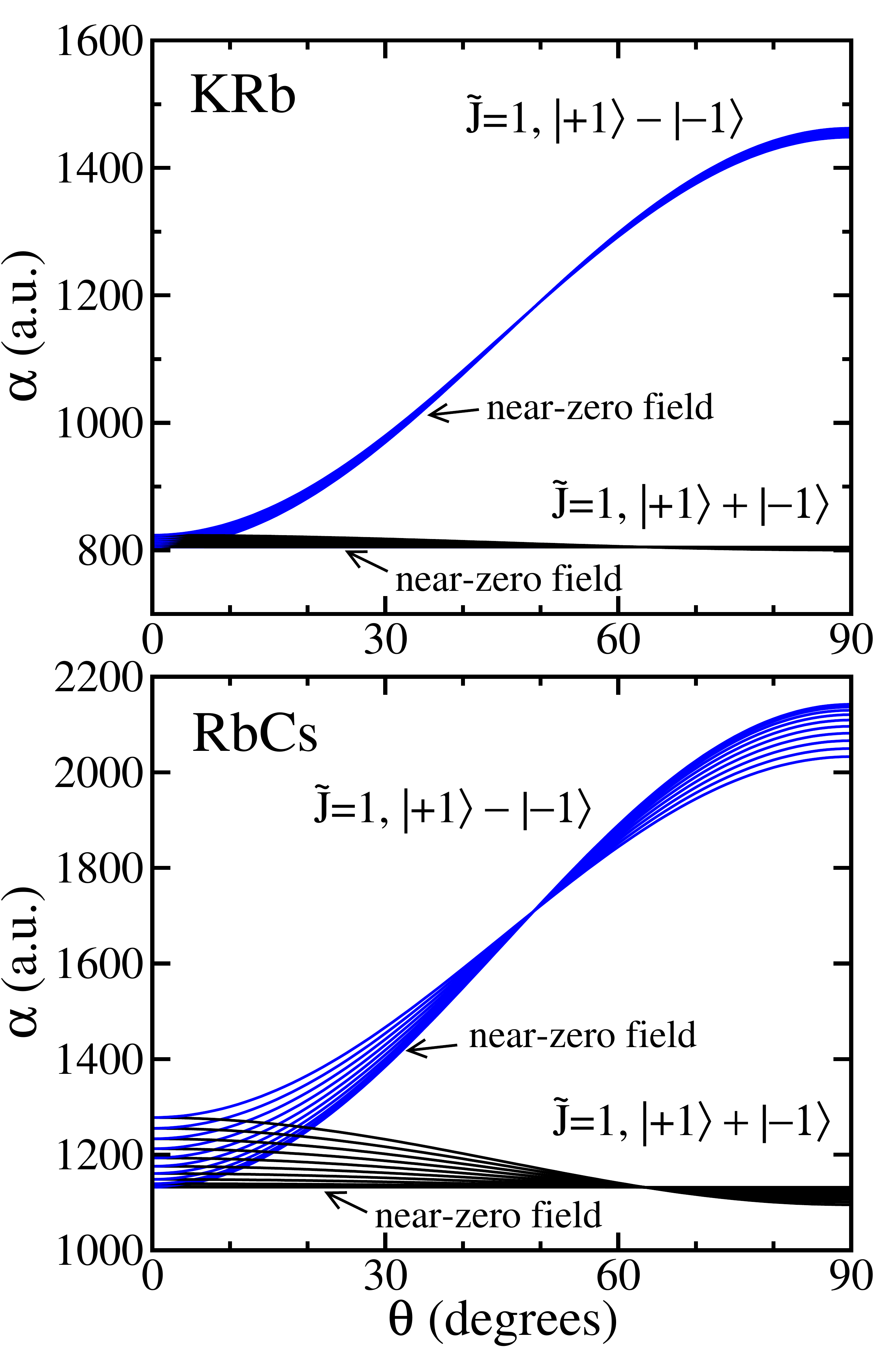}
\caption{Dependence of the dynamic polarizability on the angle between the polarization 
$\vec{\epsilon}$ of the trapping light at a wavenumber of 9174 cm$^{-1}$ and the direction of the
external electric field $\vec{\mathcal{E}}$, for different values of $\mathcal{E}$. 
The polarizability  of states with projections $M=0$  and $M=\pm 1$  of the X$^1\Sigma^+$ ground state of KRb  
and RbCs are shown in the left and right panels, respectively.
For clarity $\mathcal{E}$ ranges from 0.6 kV/cm  to 6 kV/cm in steps of 0.6 kV/cm for KRb and from 0.3 kV/cm  to 3 kV/cm  in steps of 0.3 kV/cm for 
RbCs.  At the ``magic angle'' $\theta=\theta_0=54^o$, the polarizability of the states $\tilde J=0, M=0$ and $\tilde J=1, M=0$ are 
identical and independent of the strength of the static electric field. For the $\tilde J=1, M=\pm1 $ states no magic angle exists.}
\label{magic}
\end{figure}

In future experiments one would expect to be able to change
the angle between the static and dynamic electric fields. Figure~\ref{E_vs_theta}
shows a surface plot of the dynamic polarizability as a function of 
$\mathcal{E}$ and the angle $\theta$ between $\vec{\mathcal{E}}$ and 
polarization of the trapping light $\vec{\epsilon}$, for linearly polarized light. 
The polarizability depends smoothly on both $\theta$ and $\mathcal{E}$.

Figure~\ref{magic} compares the dynamic polarizability of the states
$|\tilde J=0\rangle$ and $|\tilde J=1,M=0,\pm1\rangle$ as a function of the angle
$\theta$ of the linear polarization of the optical field relative to
the static field (such that $\hat{\epsilon} = \cos{\theta}\hat{z} +
\sin{\theta}\hat{x}$), for several static electric field strengths within
the range from 0 to 6 kV/cm for KRb and 0 to 3 kV/cm for RbCs.  All curves with $M=0$ cross at the angle $\theta =
\theta_0$ such that cos$^2\theta_0$ = 1/3, or $\theta_0 \approx 54$
degrees. We refer to this as the "magic angle" since here the AC Stark
shift is independent of the internal state of the molecule. This behavior
occurs in many contexts and is a simple consequence of the rank-2 tensor
structure of the polarizability \cite{BKDbook}.
The angular dependence of the $\tilde J=0$ level is smaller than that of the $\tilde J=1$ levels
as expected. It inherited the zero-electric field properties of the scalar $J=0$ state.
As seen in Fig.~\ref{KRb_RbCs_field_orientation}, the RbCs polarizability of the
$\tilde J=0, M=0$ and $\tilde J=1, M=0$ levels cross at a magic electric field strength of 2 kV/cm.
For Fig.~\ref{magic} this implies that the curves start to ``overlap'' when $\cal E$ is equal
or larger than this magic field value.

The polarizability operator $\alpha_{\sigma \sigma'}(\nu)$,  defined by 
Eq.~(\ref{operator_def}), is a reducible rank-two tensor operator.
Hence it can be expressed as a sum of irreducible tensor operators $\alpha^{(k)}(\nu)$ of
rank $k = 0,1,2$.  In terms of these irreducible tensor operators,
the AC Stark shift $\Delta E$ is proportional to the diagonal matrix element of  the operator $\sum_{\sigma\sigma'} \alpha_{\sigma \sigma'}(\nu)
\epsilon_{\sigma} \epsilon^*_{\sigma'}$, which can be written in the general
form
\begin{equation}
 \sum_{k=0}^{2}\sum_{q=-k}^{k} (-1)^q \alpha^{k}_q(\nu) \epsilon\epsilon^k_{-q},
\label{tensorproduct}
\end{equation}
where $q$ is a spherical tensor projection index and the
explicit forms of the irreducible spherical tensors $\alpha^{k}_q(\nu)$
and $\epsilon\epsilon^k_{q}$ are given in Refs~\cite{Bonin,Varsholovich}.

In order to derive the ``magic angle'' condition for $M=0$ states, we consider the effect
of each term in the expansion (\ref{tensorproduct}).  The term with $k=0$ corresponds
to the scalar polarizability; the operator $\alpha^{0}_0(\nu) =
\sum_{\sigma\sigma'}\alpha_{\sigma\sigma'}(\nu)\delta_{\sigma\sigma'}$ has, under our
approximations, diagonal matrix elements that are independent of $J$
(or $\tilde J$) and $M$ for all states of interest, and similarly the
quantity $\epsilon\epsilon^0_{0} \propto \vec{\epsilon}\cdot\vec{\epsilon}\,{}^*
= 1$ is independent of the polarization of the optical field.  The term
with $k=1$, corresponding to the vector polarizability, in general is
significant.  However, for the special case of linearly polarized light
where $\vec{\epsilon}$ is real, the quantities $\epsilon\epsilon^1_{q} =
(\vec{\epsilon}\times\vec{\epsilon}\,{}^*)_q$ vanish and hence the effect of
the vector polarizability is zero.  For the tensor polarizability terms
(with $k=2$), from the Wigner-Eckhart theorem only the operator component
with $q=0$ gives rise to a non-zero diagonal matrix element for $M=0$ states.  Hence the contribution
of this term is proportional to $\epsilon\epsilon^2_{0}$.  Without loss
of generality we can define the linear polarization as $\vec{\epsilon}
= \epsilon_x\hat{x} + \epsilon_z\hat{z} = \cos{\theta}\hat{z} + \sin{\theta}\hat{x}$.  
In this case $\epsilon\epsilon^2_{0} \propto \epsilon_z\epsilon_z^* - \vec{\epsilon}\cdot\vec{\epsilon}\,{}^*/3 =
\cos^2{\theta}-1/3$. Hence the contribution to $\Delta E$ due to the
tensor polarizability also vanishes for all states with $M=0$, when the optical
field is linearly polarized at the ``magic angle'' $\theta = \theta_0$.
Under this condition the only contribution to the dynamic polarizability
is from the $k=0$ scalar term, which is the
same for all states of interest.

The property  of the ``magic'' electric field discussed in
Fig.~\ref{KRb_RbCs_field_orientation} can also be understood in
terms of the tensor structure of the polarizability. 
At certain values of the applied DC electric field $\mathcal{E}$, the
rank-2 components of the polarizability, $\alpha^2_0$, of the $\tilde J=0, M=0$ and $\tilde J = 1, M = 0$ levels
becomes the same.  When this condition is met, the AC Stark shift becomes
independent of the direction of the trapping light's linear polarization.\\

Here we finish with an example of a specific implementation of a ``magic angle''
3-D lattice.  Let the lattice be formed by three orthogonal retroreflected
laser beams $a,b$, and $c$, with initial propagation directions $\hat{k}$
given by $\hat{k}_a = \hat{y}$, $\hat{k}_b = (\hat{x}+\hat{z})/\sqrt{2}$,
and $\hat{k}_c = (\hat{x}-\hat{z})/\sqrt{2}$.  These three beams each
have a different frequency $\nu$, such that $\nu_a = \nu_b -\delta_b =
\nu_c-\delta_c$; here the offset frequencies $\delta_{b,c}$ must satisfy
$\nu_{a,b,c} \gg \delta_{b,c} \gg f_{mot}$, where $f_{mot}$ is the
motional frequency of the molecules in the optical trapping potential.
The use of different frequencies (which can be generated from a single
laser by using e.g. acousto-optic modulators) in this manner
eliminates the effect of interference terms between the different laser
beams: such terms average to zero rapidly over the time of motion of the
atom, and hence can be neglected. The resulting average trap potential
is then simply the sum of the potentials due to each individual laser
beam.  Finally, the polarizations of the three beams can be chosen as
$\hat{\epsilon}_a = \sqrt{2/3}\hat{x} + \sqrt{1/3}\hat{z}$;
$\hat{\epsilon}_b = \sqrt{2/3}\hat{k}_c + \sqrt{1/3}\hat{y}$; and
$\hat{\epsilon}_c = \sqrt{2/3}\hat{k}_b + \sqrt{1/3}\hat{y}$.  In each
case, $|\hat{\epsilon}\cdot \hat{z}| = \cos{\theta_0}$.\\

\section{Acknowledgments}

This work is supported by a MURI grant of the Air Force Office of Scientific Research; by NSF; and (for DD) by DOE. SK acknowledges helpful discussions with J. Ye, D. Jin, and B. Neyenhuis.


\begin{references}

\bibitem{TrapScience08}J. Ye, H. J. Kimble, and H. Katori, Science {\bf 320}, 1734 (2008).
\bibitem{Katori}H. Katori, M. Takamoto, V. G. Pal'chikov, V. D. Ovsiannikov, Phys. Rev. Lett.
{\bf 91}, 173005 (2003).
\bibitem{Flambaum}V. V. Flambaum, V. A. Dzuba, and A. Derevianko, Phys. Rev. Lett. {\bf 101},
220801 (2008).
\bibitem{Beloy}K. Beloy, A. Derevianko, V. A. Dzuba, and V. V. Flambaum, Phys. Rev. Lett. {\bf 102},
120801 (2009).
\bibitem{Lundblad}N. Lundblad, M. Schlosser, and J. V. Porto, Phys. Rev. A {\bf 81}, 031611(R) (2010).
\bibitem{Derevianko}A. Derevianko,  Phys. Rev. A {\bf 81}, 051606(R) (2010).
\bibitem{special_issue} J. Doyle, B. Friedrich, R. V. Krems, and F. Masnow-Seeuws, Special issue on 
ulatracold polar molecules, Eur. Phys. J. D {\bf 31}, 149 (2004).
\bibitem{NJP_review} L.D. Carr, D. DeMille, R.V. Krems, Jun Ye, New J. Phys. {\bf 11}, 055049 (2009). 
\bibitem{DeMille02}D. DeMille, Phys. Rev. Lett. {\bf 88}, 067901 (2002).
\bibitem{Zhao}R. Zhao, {\it el al.} Nat. Phys. {\bf 5}, 100 (2009).
\bibitem{Zelevinsky}T. Zelevinsky, S. Kotochigova, and Jun Ye, Phys. Rev. Lett. {\bf 100}, 043201 (2008).
\bibitem{Kotochigova09}S. Kotochigova, T. Zelevinsky, and Jun Ye, Phys. Rev. A {\bf 79}, 012504 (2009).
\bibitem{Ospelkaus}S. Ospelkaus, K.-K. Ni, M. H. G. de Miranda, B. Neyenhuis, D. Wang,  S. Kotochigova,  
P. S. Julienne,  D. S. Jin, and J. Ye, Faraday Discuss. {\bf 142}, 351 (2009).
\bibitem{Micheli06} A. Micheli, G.K. Brennen, and P. Zoller, Nature Phys. {\bf 2}, 341 (2006).
\bibitem{Buchler06} H. P. B\"{u}chler, E. Demler, M. Lukin, A. Micheli, N. Prokof'ev, G. Pupillo, and P. Zoller, Phys. Rev. Lett. {\bf 98}, 060404 (2007).
\bibitem{Buchler07} H. P. B\"{u}chler, A. Micheli, and P. Zoller.  Nature Phys. {\bf 3}, 726 (2007).
\bibitem{Kotochigova1}S. Kotochigova and E. Tiesinga, Phys. Rev. A {\bf 73}, 041405(R) (2006).
\bibitem{Kotochigova2}S. Kotochigova, E. Tiesinga, P. S. Julienne, New J. Phys. {\bf 11}, 055043 (2009).
\bibitem{Kotochigova3}S. Kotochigova, New J. Phys. {\bf 12}, 073041 (2010).
\bibitem{Brink}D. M. Brink and G. R. Satchler, {\it Angular momentum}, (Clarendon Press, London, 1993).
\bibitem{Bonin} K.D. Bonin and V.V. Kresin, {\it Electric-dipole polarizabilities of atoms, molecules, and clusters}, (World Scientific, Singapore, 1997).
\bibitem{Varsholovich} D.A. Varsholovich, A.N. Moskalev, and V.K. Khersonskii, {\it Quantum theory of angular momentum: irreducible tensors, spherical harmonics, vectors coupling coefficients, 3nj symbols}, (World Scientific, Singapore, 1988).
\bibitem{Stone}A. J. Stone, {\it The theory of intermolecular forces}, (Clarendon Press, London, 1996).
\bibitem{BKDbook} D. Budker, D.F. Kimball, and D.P. DeMille, {\it Atomic Physics: an exploration through problems and solutions, 2$^\mathrm{nd}$ ed.}, (Oxford Univ. Press, Oxford, 2008).
\end{references}
\end{document}